\documentstyle[12pt]{article}

\newcommand{\EQ}{\begin{equation}}
\newcommand{\EN}{\end{equation}}

\begin{document}

\topmargin 0pt
\oddsidemargin 5mm
\newcommand{\NP}[1]{Nucl.\ Phys.\ {\bf #1}}
\newcommand{\PL}[1]{Phys.\ Lett.\ {\bf #1}}
\newcommand{\NC}[1]{Nuovo Cimento {\bf #1}}
\newcommand{\CMP}[1]{Comm.\ Math.\ Phys.\ {\bf #1}}
\newcommand{\PR}[1]{Phys.\ Rev.\ {\bf #1}}
\newcommand{\PRL}[1]{Phys.\ Rev.\ Lett.\ {\bf #1}}
\newcommand{\MPL}[1]{Mod.\ Phys.\ Lett.\ {\bf #1}}
\newcommand{\JETP}[1]{Sov.\ Phys.\ JETP {\bf #1}}
\newcommand{\TMP}[1]{Teor.\ Mat.\ Fiz.\ {\bf #1}}
     
\renewcommand{\thefootnote}{\fnsymbol{footnote}}
     
\newpage
\setcounter{page}{0}
\begin{titlepage}     
\begin{flushright}
UFSCAR-93-03
\end{flushright}
\vspace{0.5cm}
\begin{center}
\large{ Critical behaviour of integrable mixed spins chains}\\
\vspace{1cm}
\vspace{1cm}
 {\large S. R. Aladim$^{1,2}$ and  M. J.  Martins$^{1}$ } \\
\vspace{1cm}
\centerline{\em ${}^{1}$ Departamento de F\'isica, Universidade Federal de S\~ao Carlos}
\centerline{\em S\~ao Carlos, C.P. 676, 13560, S\~ao Paulo, Brasil}
\centerline{\em ${}^{2}$ Departamento de F\'isica e Ci\^encias de Materiais,  }
\centerline{\em  Instituto de F\'isica e Qu\'imica de S\~ao Carlos}
\centerline{\em S\~ao Carlos, 13560, S\~ao Paulo, Brasil}
\vspace{1.2cm}   
\end{center} 
\begin{abstract}
We construct a mixed spin 1/2 and $S$  integrable model and investigate its
finite size properties. For a certain
conformal invariant mixed spin system the central charge can be decomposed in terms of the
conformal anomaly of two single integrable models of spin 1/2 and spin $(S-1/2)$. We also compute the ground state
energy and the sound velocity in the thermodynamic limit.
\end{abstract}
\vspace{.2cm}
\vspace{.2cm}
\centerline{Published in J.Phys.A:Math.Gen.26 (1993) L529 }
\end{titlepage}

\renewcommand{\thefootnote}{\arabic{footnote}}
\setcounter{footnote}{0}

\newpage

Integrable magnetic spin chains provide important examples of systems which can be derived from
the so-called Yang-Baxter algebra \cite{YB}. A well known model is the 
isotropic spin 1/2 Heisenberg \cite{BE}
chain and its generalization for arbitrary 
spin $S$ \cite{TA,BA}. Another interesting example is the Heisenberg
model in presence of an impurity of spin $S$ \cite{NH,SO}. In 
such model one of the local vertex weight acts
on a pair of asymmetric vector spaces which 
is defined  by the local states on the horizontal and vertical lines of a two 
dimensional lattice. More recently, a general discussion concerning the construction of mixed vertex models
has been presented by de Vega and Woynorovich \cite{VW}. For instance, they have studied several properties
of the thermodynamic limit of an alternating anisotropic chain of spins 1/2 and 1. However,
 it is still to be 
investigated the finite size effects in these mixed spins models as well as their class of
universality for conformally invariant systems. Following the approach of ref. \cite{VW}
we construct an isotropic alternating spin 1/2 and $S$ chain. We focus our attention in the analysis of
the finite size behaviour of the ground state on a line of length $L$. A conformally invariant mixed
system can be defined and its conformal anomaly is computed by analysing the finite size corrections
for the ground state energy and by the thermodynamic Bethe ansatz. Several 
useful quantities as the ground
state energy and the sound velocity are also computed.

The construction of the transfer matrix of the mixed spins $\sigma$ and $S$ model 
is based on the local vertex  $ R_{S,j}^{\sigma}(\lambda)$ which is a matrix
in the auxiliary space $V_{\sigma}$ and its matrix elements are operators of spin $S$ acting
on the Hilbert state space at site $j$. In the case of an auxiliary space of spin 1/2,  $R_{S,j}^{1/2}(\lambda)$
\cite{KRS} is given by
\EQ
R_{S,j}^{1/2}(\lambda) = \left( \begin{array}{cc}
  \lambda  S^{0}  +  i S^{3}   &   iS^{-}  \\
  i S^{+}   &   \lambda  S^{0} - i S^{3}  
\end{array}
\right) ~~ S_j^{\pm}= S_j^1 \pm i S_j^2 
\EN
where $S_j^a$, $a=1,2,3$ are spin $S$ operators and $S^{0}$ is a $(2S+1)\times (2S+1)$  
identity matrix.

In the case of an alternating spin 1/2 and $S$ mixed model the set of
 commuting transfer matrix $ T_{1/2,S}(\lambda) $ 
 assumes the
following form
\EQ
T_{1/2,S}(\lambda)=Tr_{V_{1/2}}(\tau_{1/2,S}(\lambda)),~~ \tau_{1/2,S}(\lambda)=R_{1/2,L}^{1/2}(\lambda)
R_{S,L-1}^{1/2}(\lambda) \cdots R_{S,1}^{1/2}(\lambda)
\EN
where $\tau_{1/2,S}$ is a (2x2) matrix in the auxiliary space $V_{1/2}$ 
denominated the monodromy matrix. Here 
we impose periodic boundary conditions and  the length $L$ is an even number.

The associated one dimensional quantum Hamiltonian which commutes with $T_{1/2,S}$ is defined by
$H_{1/2,S}= i J \frac{d}{d \lambda} \log(T_{1/2,S}(\lambda))|_{\lambda=i/2}$ and has the following
expression
\EQ
H_{1/2,S}= \tilde{J}  \left[ \sum_{n=even}^{L } \sum_{i,j=0}^{3}
 \sigma^i_{n-1}\{S^i_n,S^j_n\}\sigma^j_{n+1} + 
\sum_{i=1}^3(\frac{1}{4} - S(S+1)) \sigma^i_{n-1} \sigma^i_{n+1} \right] 
- \frac{\tilde{J} L(2S+3)^2}{8}
\label{eq:hamiltonian}
\EN
where  $\sigma^i$, $i=1,2,3$  are Pauli matrix elements, $\sigma^0 $ is the identity matrix, and
$\tilde{J} = 2J/(2S+1)^2 $ .
In this paper we are interested in the antiferromagnetic $(J > 0)$ 
regime of (\ref{eq:hamiltonian}) and let us assume, for the sake of simplicity, $J=1$.

Similarly to the usual spin 1/2 Heisenberg chain, the Hamiltonian~(\ref{eq:hamiltonian})
 can be diagonalized by the
quantum inverse scattering method \cite{QI}. The  eigenenergies are parametrized in terms of the complex
numbers $\lambda_j$
\EQ
E_{1/2,S}= -\sum_{j=1}^M \frac{1}{\lambda_j^2 +1/4} 
\EN
where $\lambda_j$ satisfy the so-called Bethe ansatz equation
\EQ
{\left( \frac{\lambda_j -i/2}{\lambda_j +i/2} \right)}^{L/2}
{\left( \frac{\lambda_j -iS}{\lambda_j +iS} \right)}^{L/2}= -\prod_{l=1}^M 
\frac{\lambda_j-\lambda_l-i}{\lambda_j-\lambda_l+i}
\label{eq:BetheAns}
\EN
where $M=\frac{L}{2}(S+1/2)-r$, and $r$ labels the disjoint sectors of the eigenvalues of the total
spin operator $\sum_{j=odd}^M (\sigma_j)+ \sum_{j=even}^M (S_j)$.

In order to find the structure of the numbers $\lambda_j$ for finite size systems 
we numerically solve Eqs.(4,5) and compare them with the exact diagonalization of the Hamiltonian (3). 
In Fig.1(a,b) we show the picture of the ground state for $L=6,8,10$ and $S=1,3/2$. In Table 1 we present the
respective values of the complex parameter $\lambda_j$. For large enough $L~(L \geq 8)$ the numbers
$\lambda_j$ cluster in two distint sets of roots. One of them consists of real numbers~(full circles) and the other of
complex structures~(crosses) $\lambda_j^{\alpha}$ which 
in the asymptotic  limit $L \rightarrow \infty$ are 
called 2S-strings
\EQ
\lambda_j^{\alpha}= \xi_j +\frac{i}{2}(2S+1-2 \alpha),~~ \alpha=1,2, ..., 2S
\EN
where $\xi_j$ is a real number denominated center of the 2S-string.

Taking into account the structure discussed above, the ground state energy per particle $e_{\infty}^{1/2,S}$
can be calculated and is given by
\EQ
e_{\infty}^{1/2,s}= -\ln(2) -\frac{1}{2} \left( \psi(\frac{2S+3}{4})-\psi(\frac{2S+1}{4}) \right)
\EN
where $\psi(x)$ is the Euler psi function.

Analogously, it is possible to define the transfer matrix $T_{S,1/2}(\lambda)$
of a mixed spin $S$ and 1/2 model which commutes with $T_{1/2,S}(\lambda)$. The transfer matrix $T_{S,1/2}(\lambda)$ has the following
expression

\EQ
T_{S,1/2}(\lambda)=Tr_{V_{S}}(\tau_{S,1/2}(\lambda)),~~ \tau_{S,1/2}(\lambda)=R_{1/2,L}^{S}(\lambda)
R_{S,L-1}^{S}(\lambda)....R_{S,1}^{S}(\lambda)
\EN

The commutativity between $T_{1/2,S}(\lambda)$ and $T_{S,1/2}(\lambda)$ derives from the 
Yang-Baxter relations satisfied by the local vertices $R_{S,j}^{\sigma}$ \cite{KRS}. Interesting enough, this
permits us to define a $rotational$ invariant mixed vertex model by formally multiplying the two transfer
matrix \cite{VW}
\EQ
T^{sym}=T_{1/2,S} T_{S,1/2}
\EN

Due to commutativity $[T_{1/2,S}(\lambda),T_{S,1/2} \lambda^{'}] = 0 $, the eigenenergies of the one-dimensional
Hamiltonian associated to $T^{sym}$ are parametrized by the same Bethe equations, namely Eq.\ref{eq:BetheAns}. However,
the expression for the total energy for a given set of numbers $\lambda_j$ is

\EQ
E^{sym}= -\sum_{j=1}^M \frac{1}{\lambda_j^2 +1/4} 
- \sum_{j=1}^M \frac{2S}{\lambda_j^2 +S^2} 
\label{eq:energia}
\EN

The model defined by  Eqs.(\ref{eq:BetheAns}, \ref{eq:energia}) possesses 
all features of being conformally invariant, i.e, short-range
interaction, translational 
and $rotational$ symmetry and gapless low-lying excitations in the spectrum. Indeed
for low (total) momenta $p$ \footnote{We remark here that the correct
momentum operator is half of that used in ref.[7]~(see eq.(2.35)). This
implies that the sound velocity is double~($v_s=2 \pi)$ of that found
previously in ref.[7]~($v_s=\pi$).},the dispersion relation is linear in $p$
\EQ
\epsilon(p) \sim v_s p
\EN
where $v_s=2\pi$ is the sound velocity.

The class of universality of this model can be determined by exploiting a set of important relations \cite{CA}
between the eigenspectrum of finite lattice systems. In particular the conformal anomaly is related to the
ground state energy $E_0^{sym}(L)$ by \cite{CA1,AF}
\EQ
E_0^{sym}(L)/L = e_{\infty}^{sym} -\frac{\pi c v_s}{6 L^2}
\label{eq:Einfty}
\EN
where $c$ is the central charge and $e_{\infty}^{sym}$ is the ground state per particle
\EQ
e_{\infty}^{sym}= e_{\infty}^{1/2,S} -\frac{1}{2} \left[ \psi(\frac{2S+3}{4}) 
-\psi(\frac{2S+1}{4}) + \psi(\frac{2S+1}{2})
-\psi(1/2) \right] 
\EN

In Table 2 we present our estimatives for 
the central charge $c$ of Eq.(\ref{eq:Einfty}) for $S=1,3/2$. Our numerical result
predicts a conformal anomaly $c=2.01(2)(S=1)$ and $c=2.500(6)(S=3/2)$. We notice that for $S=1$ the extrapolation
is less precise compared to $S=3/2$. This is due to the fact that the bulk of the complex part of
2-string zeros are next to $\pm i/2$. As a consequence, we can 
apply an analytical technique developed
by de Vega and Woynorovich \cite{VW1} and we find the exact value $c=2$ for $S=1$.

Another efficient method to compute the conformal anomaly is by analysing the low temperature behaviour
of the associated free energy. For a critical system at low temperature, the free energy per particle has
the following asymptotic behaviour \cite{CA1,AF}
\EQ
F(T)/L = e_{\infty} -\frac{\pi c T^2}{6 v_s}
\label{eq:Free0}
\EN

In the case of integrable one dimensional spin chains the thermodynamic properties can be studied by using
the thermodynamic Bethe ansatz method. In this approach, the free energy is given in terms of 
variables denominated pseudoenergies and its minimization yields a set of integral equations for these
parameters. In order to obtain such equations for the mixed spin model defined by Eqs.(\ref{eq:BetheAns}, \ref{eq:energia}) we follow 
refs. \cite{TA,BA} and here we give only the final results. The free energy at temperature T is given by
\EQ
F^{sym}(T)/L =e_{\infty}^{sym} -\frac{T}{4 \pi} \int_{-\infty}^{+\infty} p(\lambda) \left ( \ln(1+e^{\epsilon_1(\lambda)/T} ) +\ln(1+ e^{\epsilon_{2S}(\lambda)/T} \right )
\label{eq:Free}
\EN
where the pseudoenergies $\epsilon_a(\lambda), a=1,2,3,\ldots$ satisfy the following TBA equations
\EQ
\epsilon_n(\lambda)/T =p(\lambda)* \left( \ln(1+ e^{\epsilon_{n+1}/T}) +\ln(1+e^{\epsilon_{n-1}/T}) \right )
+\frac{p(\lambda)} {T}( \delta_{n,1} + \delta_{n,2S})
\label{eq:Pseudo}
\EN
where $  \epsilon_{0}(\lambda)=0$, $f*g(x)$ denotes the convolution $\frac{1}{2 \pi} \int_{-\infty}^{+\infty} f(x)g(x-y) dy$ and 
$p(\lambda)=\frac{\pi}{\cosh(\pi \lambda)}$.

The advantage of Eqs.(\ref{eq:Free}, \ref{eq:Pseudo}) is that they allow us to exactly estimate the low temperature behaviour of the free 
energy. Using the standard procedure~( see  e.g. \cite{TAK,BA}) to compute the leading behaviour of $F(T)$, and
some dilogarithm identities \cite {KI} we find the  following result
\EQ
F(T)/L = e_{\infty} - \frac{4S-1}{6(2S+1)} T^2
\label{eq:Free2}
\EN

Comparing Eqs.(\ref{eq:Free2}) and (\ref{eq:Free0}) the value of the central charge is given by
\EQ
c= \frac{2(4S-1)}{2S+1}
\label{eq:central}
\EN

For $S=1,3/2$ we obtain $c=2,2.5$ which are consistent with our numerical findings of Table 2. Interesting
enough, Eq.(\ref{eq:central}) can be decomposed in terms of the central charges of the integrable Heisenberg chains of
spins 1/2~($c=1$) and $S-1/2$~($c=\frac{3(2S-1)}{2S+1})$. In this sense the effect
of the interaction between spins 1/2 and $S$ is the ``reduction'' of the critical behaviour of a spin $S$ to
$S-1/2$ \footnote{ We remark that $ J=S-1/2$ is the smallest possible value in the   addition of two "angular momenta", $S$ and $1/2$.}. Taking into account this last observation it is easy to conjecture the conformal anomaly of 
a mixed spin $S$ and $S^{'}$. In general, we can assume $ S^{'}> S$ and the expected
critical behaviour is given by the following conformal anomaly 
\EQ
c= \frac{3S}{S+1} + \frac{3(S^{'}-S)}{S^{'}-S+1}
\label{eq:Central2}
\EN

We believe that result (\ref{eq:Central2}) is the first step toward the understanding of the composition
of the operator content of these mixed spins models. However, 
it is still to be investigated how the primary 
fields are composed in these systems. We hope to report on this problem in a future publication.
\section*{Acknowledgements}
S.R. Aladim is indebted to F.C. Alcaraz for suggesting this problem and his 
collaboration in an early stage of this work  and to R. K\"oberle for discussions.
We thanks A. Malvezzi for his help with numerical diagonalization. The work of M.J. Martins was partially supported
by CNPq~(Brazilian agency).

\newpage
\centerline{ \bf Figure Captions}
\vspace{0.5cm}
Fig.1(a,b) The ground state configurations of the numbers $\lambda_j$ for
$L=6,8,10$ and $S=1,3/2$.
\vspace{1.cm}\\
\centerline{ \bf Table Captions}
\vspace{0.5cm}\\
Table 1. The numerical values of the numbers $\lambda_j$ of the Bethe ansatz
equation (5) corresponding to the ground state for $L=6,8,10$ and $S=1,3/2$.\\

Table 2. The estimatives of the conformal anomaly of Eq.(12) for $S=1$ and $S=3/2$ 

\newpage
\begin{center}

{\bf Table 1}\\
\vspace{0.5cm}
\begin{tabular}{|l|l|l|} \hline
L & $ S=1$ & $S=3/2$ \\ \hline \hline
6 & $ 0.316 194 \pm i 0.508 074 $ & $\pm i 1.5$, $ \pm i 0.5$ \\  
  & $-0.494 077$, $ 0.228 965$ & $\pm 0.358 639$ \\ \hline
8 & $\pm 0.308 362 \pm i 0.503 329$ & $\pm 0.302 413 \pm i 1.060 323$ \\
  &  $\pm 0.308 005 $ & $ \pm 0.436 072$, $ \pm 0.201 077$ \\ \hline
 & $\pm 0.535 111 \pm i 0.5101 82$ & $\pm 0.302 415 \pm i 1.060 320$ \\ 
10  & $\pm 0.444 828$, $ 0 $ & $\pm 0.436 069$, $\pm i 0.5$  \\ 
  & & $\pm 0.201 079$ \\ \hline
\end{tabular}

\vspace{2.0cm}
{\bf Table 2}\\
\vspace{0.5cm}
\begin{tabular}{|l|l|l|} \hline
L & $ S=1$ & $S=3/2$ \\ \hline \hline
8 & 2.210 923 & 2.839 364 \\ \hline
16 & 2.061 249 & 2.602 543 \\ \hline
24 & 2.041 930 & 2.556 301 \\ \hline
32 & 2.020 828 & 2.538 530 \\ \hline
40 & 2.017 593 & 2.529 503 \\ \hline
48 & 2.014 692 & 2.524 142 \\ \hline
56 & 2.012 134 & 2.520 624 \\ \hline
Extrapolated & 2.01(1) & 2.500(6) \\ \hline
\end{tabular}
\end{center}

\newpage

\unitlength=1.00mm
\special{em:linewidth 0.4pt}
\linethickness{0.4pt}
\begin{picture}(181.40,176.82)(30.00,0.00)
\put(4.32,53.92){\line(1,0){50.00}}
\put(29.42,5.88){\line(0,1){98.04}}
\put(36.32,53.92){\makebox(0,0)[cc]{$\bullet$}}
\put(22.26,53.92){\makebox(0,0)[cc]{$\bullet$}}
\put(34.12,67.17){\makebox(0,0)[cc]{$0.5$}}
\put(34.12,39.88){\makebox(0,0)[cc]{$-0.5$}}
\put(28.32,39.88){\line(1,0){2.00}}
\put(28.32,67.17){\line(1,0){2.00}}
\put(34.12,92.94){\makebox(0,0)[cc]{$1.5$}}
\put(34.12,14.90){\makebox(0,0)[cc]{$-1.5$}}
\put(28.32,93.06){\line(1,0){2.00}}
\put(28.32,14.78){\line(1,0){2.00}}
\put(29.42,67.17){\makebox(0,0)[cc]{$\times$}}
\put(29.42,92.94){\makebox(0,0)[cc]{$\times$}}
\put(29.42,39.88){\makebox(0,0)[cc]{$\times$}}
\put(29.42,14.90){\makebox(0,0)[cc]{$\times$}}
\put(41.18,100.00){\makebox(0,0)[cc]{$L=6$}}
\put(21.18,110.00){\makebox(0,0)[cc]{\large (b)}}
\
\
\put(69.32,53.92){\line(1,0){50.00}}
\put(94.12,9.02){\line(0,1){89.80}}
\put(100.22,53.92){\makebox(0,0)[cc]{$\bullet$}}
\put(88.42,53.92){\makebox(0,0)[cc]{$\bullet$}}
\put(106.79,80.94){\makebox(0,0)[cc]{$\times$}}
\put(106.79,25.94){\makebox(0,0)[cc]{$\times$}}
\put(83.24,25.94){\makebox(0,0)[cc]{$\times$}}
\put(83.24,80.94){\makebox(0,0)[cc]{$\times$}}
\put(100.45,80.12){\makebox(0,0)[cc]{$1.0$}}
\put(100.45,27.72){\makebox(0,0)[cc]{$-1.0$}}
\put(93.32,80.12){\line(1,0){2.00}}
\put(93.32,27.72){\line(1,0){2.00}}
\put(77.26,53.72){\makebox(0,0)[cc]{$\times$}}
\put(111.38,53.72){\makebox(0,0)[cc]{$\times$}}
\put(102.36,100.00){\makebox(0,0)[cc]{$L=8$}}
\
\
\put(156.48,9.02){\line(0,1){89.80}}
\put(131.40,53.92){\line(1,0){50.00}}
\put(166.26,53.92){\makebox(0,0)[cc]{$\bullet$}}
\put(146.54,53.92){\makebox(0,0)[cc]{$\bullet$}}
\put(169.26,80.94){\makebox(0,0)[cc]{$\times$}}
\put(169.26,26.72){\makebox(0,0)[cc]{$\times$}}
\put(144.54,26.72){\makebox(0,0)[cc]{$\times$}}
\put(144.54,80.94){\makebox(0,0)[cc]{$\times$}}
\put(162.53,67.17){\makebox(0,0)[cc]{$0.5$}}
\put(162.53,39.88){\makebox(0,0)[cc]{$-0.5$}}
\put(155.40,67.17){\line(1,0){2.00}}
\put(155.40,39.88){\line(1,0){2.00}}
\put(156.40,53.92){\makebox(0,0)[cc]{$\bullet$}}
\put(136.48,53.72){\makebox(0,0)[cc]{$\times$}}
\put(177.26,53.72){\makebox(0,0)[cc]{$\times$}}
\put(162.80,80.12){\makebox(0,0)[cc]{$1.0$}}
\put(155.67,80.12){\line(1,0){2.00}}
\put(162.80,27.72){\makebox(0,0)[cc]{$-1.0$}}
\put(155.67,27.72){\line(1,0){2.00}}
\put(168.24,98.82){\makebox(0,0)[cc]{$L=10$}}
\put(156.48,67.06){\makebox(0,0)[cc]{$\times$}}
\put(156.48,40.00){\makebox(0,0)[cc]{$\times$}}
\
\put(4.32,144.82){\line(1,0){50.00}}
\put(29.32,118.82){\line(0,1){51.00}}
\put(36.32,144.82){\makebox(0,0)[cc]{$\bullet$}}
\put(9.32,144.82){\makebox(0,0)[cc]{$\bullet$}}
\put(28.32,128.82){\line(1,0){2.00}}
\put(28.32,160.82){\line(1,0){2.00}}
\put(69.32,144.82){\line(1,0){50.00}}
\put(94.32,118.82){\line(0,1){51.00}}
\put(105.32,144.82){\makebox(0,0)[cc]{$\bullet$}}
\put(83.32,144.82){\makebox(0,0)[cc]{$\bullet$}}
\put(110.32,162.82){\makebox(0,0)[cc]{$\times$}}
\put(110.32,126.82){\makebox(0,0)[cc]{$\times$}}
\put(79.32,126.82){\makebox(0,0)[cc]{$\times$}}
\put(79.32,162.82){\makebox(0,0)[cc]{$\times$}}
\put(98.49,160.82){\makebox(0,0)[cc]{$0.5$}}
\put(98.49,128.82){\makebox(0,0)[cc]{$-0.5$}}
\put(93.32,160.82){\line(1,0){2.00}}
\put(93.32,128.82){\line(1,0){2.00}}
\put(131.40,144.82){\line(1,0){50.00}}
\put(156.40,118.82){\line(0,1){51.00}}
\put(169.40,144.82){\makebox(0,0)[cc]{$\bullet$}}
\put(143.40,144.82){\makebox(0,0)[cc]{$\bullet$}}
\put(172.40,162.82){\makebox(0,0)[cc]{$\times$}}
\put(172.40,126.82){\makebox(0,0)[cc]{$\times$}}
\put(141.40,126.82){\makebox(0,0)[cc]{$\times$}}
\put(141.40,162.82){\makebox(0,0)[cc]{$\times$}}
\put(160.57,160.82){\makebox(0,0)[cc]{$0.5$}}
\put(160.57,128.82){\makebox(0,0)[cc]{$-0.5$}}
\put(155.40,160.82){\line(1,0){2.00}}
\put(155.40,128.82){\line(1,0){2.00}}
\put(156.40,144.82){\makebox(0,0)[cc]{$\bullet$}}
\put(29.42,176.82){\makebox(0,0)[cc]{$L=6$}}
\put(94.12,176.82){\makebox(0,0)[cc]{$L=8$}}
\put(156.48,176.82){\makebox(0,0)[cc]{$L=10$}}
\put(41.18,127.06){\makebox(0,0)[cc]{$\times$}}
\put(41.18,161.96){\makebox(0,0)[cc]{$\times$}}
\put(23.14,160.00){\makebox(0,0)[cc]{$0.5$}}
\put(23.14,127.84){\makebox(0,0)[cc]{$-0.5$}}
\put(19.42,178.82){\makebox(0,0)[cc]{\large(a)}}
\end{picture}
\centerline{\bf Figure 1(a,b)}

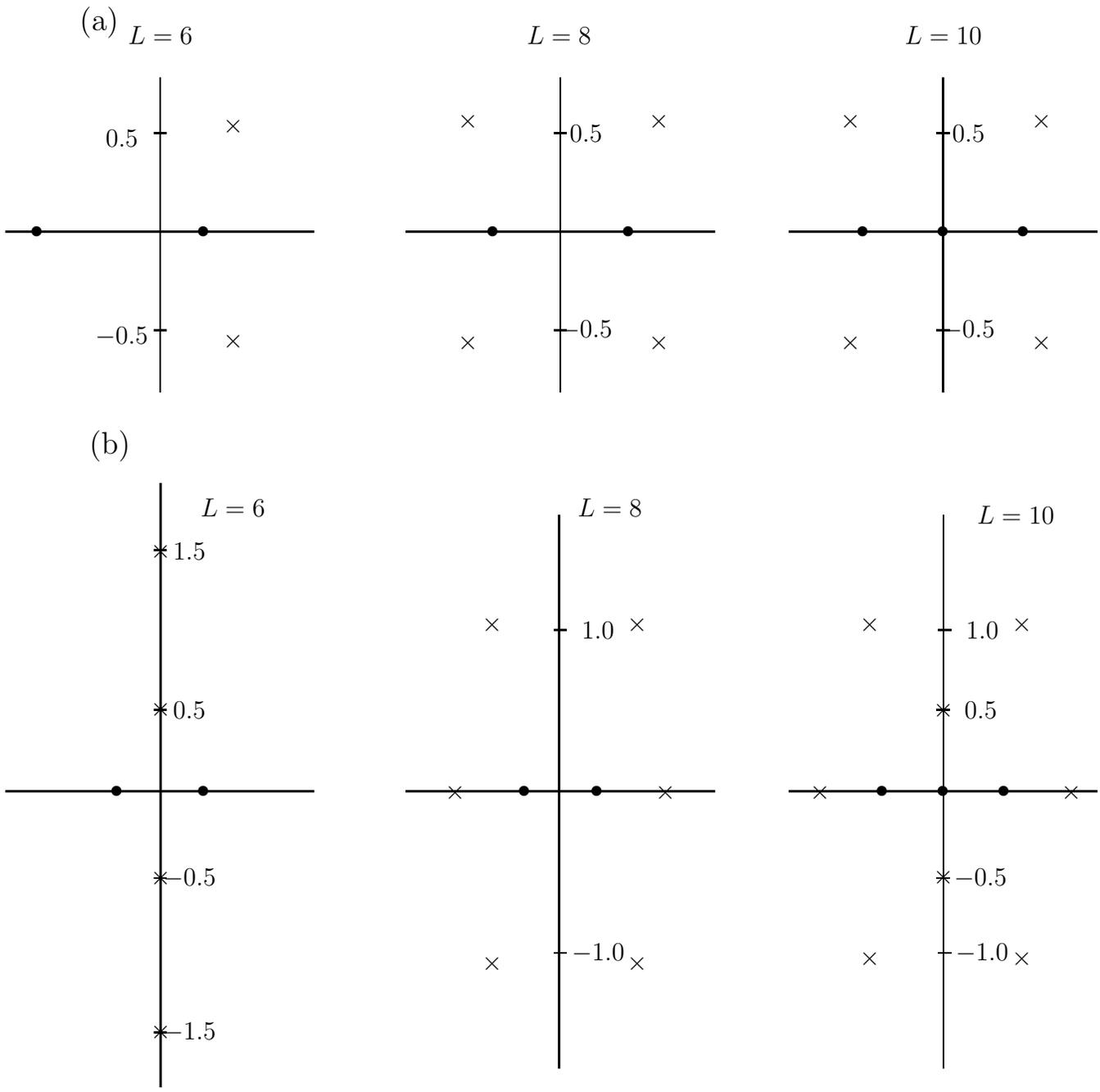
\end{document}